\documentclass[11pt,letterpaper]{article}
\usepackage[margin=1in]{geometry}
\usepackage[T1]{fontenc}
\usepackage[utf8]{inputenc}
\usepackage{lmodern}
\usepackage{amsmath,amssymb,amsthm}
\usepackage{xcolor}
\usepackage[hidelinks]{hyperref}

\newtheorem{theorem}{Theorem}[section]
\newtheorem{lemma}{Lemma}[section]
\newtheorem{problem}{Problem}[section]

\title{Hardness of algorithmic Kadison-Singer problem from approximate\\
hypergraph colouring}
\author{Danny Vagnozzi \thanks{This author is supported the UK EPSRC grant EP/X033201/1.}\\
Durham University\\
\texttt{danny.vagnozzi@durham.ac.uk}}
\date{November 20, 2026}

\begin{document}
\maketitle

\begin{abstract}
In 2015, Marcus, Spielman and Srivastava published a solution of a strong form of a conjecture by Weaver implying a positive answer to the Kadison-Singer problem - an open question of interest to several fields of pure and applied mathematics. Specifically, they proved that for all $\alpha>0$ and finite lists of vectors whose Euclidean norm is at most $\sqrt\alpha$ and whose outer products sum to the identity matrix, there exists a signed sum of these outer products whose operator norm is at most $\sqrt{8\alpha}+2\alpha$. Since then, significant effort has been put into understanding the algorithmic side of the Kadison-Singer problem; that is, the task of actually finding a signing resulting in a matrix with operator norm at most $c\sqrt\alpha$ for some fixed constant $c$. The first hardness results for this problem were proven independently by Spielman and Zhang [APPROX'22] and by Jourdan, Macgregor and Sun [ISAAC'23].

In this short note, we prove that there exist values of $c$ arbitrarily close to $1$ such that it is NP-hard to distinguish indexed families of vectors for which there is a signed sum with operator norm equal to $0$ from those for which every signed sum has operator norm at least $c\sqrt\alpha$. This improves on the previous best bound by Spielman and Zhang. Our proof borrows significant ideas from the above mentioned hardness results, by adapting them to a reduction from the problem of approximating hypergraph colouring in the sense of promise constraint satisfaction problems.
\end{abstract}

\section{Introduction}

The \emph{Kadison-Singer problem} has been studied in various equivalent guises for decades before being formalised in 1959 \cite{KadisonSinger1959}. The problem was shown to be equivalent to a number of important open problems in fields stretching from pure mathematics to engineering. Whilst the original formulation is in the language of abstract operator algebra, Weaver \cite{Weaver2004} showed that it is implied by a more elementary linear algebraic statement concerning a `discrepancy function'. A seminal result by Marcus, Spielman and Srivastava \cite{MarcusSpielmanSrivastava2015} proved the following stronger form of this statement, thus solving the Kadison-Singer problem.

\begin{theorem}\label{thm:mss}
Fix an indexed family $\mathcal V=(v_1,\ldots,v_m)$ of vectors $v_i\in\mathbb C^d$ and a constant $\alpha>0$ such that $\|v_i\|_2^2\le\alpha$ for all $i\in[m]$. Assume that $\sum_{i\in[m]}v_iv_i^\dagger=I_d$. For a map $\sigma:\mathcal V\to\{\pm1\}$, define
\[
W_\sigma(\mathcal V)=\left\|\sum_{i\in[m]}\sigma(v_i)v_iv_i^\dagger\right\|_2.
\]
Let $W(\mathcal V)=\min_\sigma W_\sigma(\mathcal V)$. Then there is a universal constant $c$ for which $W(\mathcal V)\le c\sqrt\alpha$.
\end{theorem}

The optimal value for this constant is unknown. Recent work announced by Song and Yue \cite{SongYue2026TwoDeviations} bounds the optimal $c$ by $2\le c\le2.176$. For $0<\alpha\le1/4$, it has been shown that $W(\mathcal V)\le\sqrt{8\alpha(1-2\alpha)}$ \cite{BownikCasazzaMarcusSpeegle2019}.

Of interest to theoretical computer science is whether, given a finite indexed family $\mathcal V$ in $\mathbb C^d$, one can efficiently find a signing $\sigma:\mathcal V\to\{\pm1\}$ so that $W_\sigma(\mathcal V)\le c\sqrt\alpha$ for some fixed constant $c$. While it is unclear from the existential proof in Theorem~\ref{thm:mss} if such an algorithm exists, the task of finding a suitable $\sigma$ is closely related to the problem of constructing unweighted spectral sparsifiers \cite{batson2009twice}. In line with recent work by Jourdan, Macgregor and Sun \cite{JourdanMacgregorSun2023}, we call the following the decision version of the algorithmic Kadison-Singer problem and denote it by $\mathsf{KS}_2(c)$.

\begin{problem}[Algorithmic Kadison-Singer problem (decision version)]\label{prob:ks}
Fix a constant $c>0$. Given as input a positive rational $\alpha$ and an indexed family $\mathcal V=(v_1,\ldots,v_m)$ such that $v_i\in\mathbb C^d$, $\|v_i\|_2^2\le\alpha$ for all $i\in[m]$ and
\[
\sum_{i\in[m]}v_iv_i^\dagger=I_d,
\]
decide whether there is a signing $\sigma:\mathcal V\to\{\pm1\}$ such that
\[
\left\|\sum_{i\in[m]}\sigma(v_i)v_iv_i^\dagger\right\|_2\le c\sqrt\alpha.
\]
\end{problem}

The search version consists in actually finding the map $\sigma$, or reporting that no such map exists. A polynomial-time algorithm for this search problem would therefore decide the decision version with the corresponding parameters.

There has been significant progress on the study of the complexity of $\mathsf{KS}_2(c)$ in recent years. The first hardness results were proven by Spielman and Zhang and independently by Jourdan, Macgregor and Sun \cite{SpielmanZhang2022,JourdanMacgregorSun2023}. With very similar constructions, both papers show that, even if we fix $\alpha=\frac14$, $\mathsf{KS}_2(c)$ is NP-hard for any fixed $0<c<1/2$ and $0<c<1/(2\sqrt2)$, respectively. Spielman and Zhang further showed that there is a constant $c$ such that the problem is NP-hard for arbitrarily small fixed $\alpha$. Recent results have announced deterministic polynomial time algorithms for values of $c=35$ \cite{Kathuria2026Weaver} and $c=13$ \cite{EzeunalaJiang2026}. Other interesting algorithmic results were announced by Jourdan et al.: in \cite{JourdanMacgregorSun2024Dense}, the authors provided a deterministic algorithm achieving $W_\sigma(\mathcal V)\le1/2$ for equal-norm inputs under the condition $m\ge49d^2$ (i.e. the input is `dense'), whereas in \cite{JourdanMacgregorSun2023} they give a randomised algorithm for approximating a solution, with a running time exponential in $d$ that is quasipolynomial in $m$ in low-dimensional regimes. One of the motivating questions for these results is the following:

\begin{quote}
What is the supremum of the constants $c_K>0$ such that $\mathsf{KS}_2(c_K)$ is NP-hard?
\end{quote}

In this short note, we show the following.

\begin{theorem}\label{thm:main}
For every $\epsilon\in(0,1)$ there exists $\delta\in(0,1)$ such that $1-\delta>\epsilon$ and $\mathrm{KS}_2(1-\delta)$ is NP-hard.
\end{theorem}

Our proof techniques are very similar to those of \cite{SpielmanZhang2022,JourdanMacgregorSun2023}, though we generalise the construction in their reduction proofs to $r$-uniform hypergraphs for arbitrarily large values of $r$. This is so that we can use the problem of approximating almost balanced hypergraph $2$-colouring by $3$-colouring as source of hardness \cite{GuruswamiLee2018}. This is the problem of distinguishing $r$-uniform hypergraphs that can be almost balanced $2$-coloured (in a well defined way) from $r$-uniform hypergraphs that cannot be $3$-coloured, and was shown to be NP-hard by Guruswami and Lee \cite{GuruswamiLee2018}. Specifically, for a fixed value of $r$, our reduction maps $r$-uniform hypergraphs $\mathbf H$ to instances $\mathcal V_{\mathbf H}$ of $\mathrm{KS}_2(c)$ (for a suitable $c$ depending on $r$) such that:
\begin{enumerate}
\item $W(\mathcal V_{\mathbf H})=0$ whenever $\mathbf H$ is almost balanced $2$-colourable.
\item $W(\mathcal V_{\mathbf H})\ge\beta_r>c\sqrt{\alpha_r}$ whenever $\mathbf H$ is not $3$-colourable.
\end{enumerate}
Here $\alpha_r$ and $\beta_r$ are specified in Theorem~\ref{thm:fixedarity}.

Another way our proof differs from \cite{SpielmanZhang2022,JourdanMacgregorSun2023} is the technique we use to achieve the second point: whilst for the authors in \cite{SpielmanZhang2022,JourdanMacgregorSun2023} it was sufficient to convert a signing $\sigma$ with sufficiently small $W_\sigma(\mathcal V_{\mathbf H})$ to a solution of a Boolean constraint satisfaction problem, we use a more intricate rounding to convert it to a $3$-colouring of a hypergraph.

\section{Approximate hypergraph colouring problems}

Before delving into the proof of Theorem~\ref{thm:main}, we briefly outline some notions on approximate hypergraph colouring problems, which is where our hardness reduction starts.

\paragraph{Hypergraphs.}
An \emph{$r$-uniform hypergraph} is a pair $\mathbf H=(V,E)$ where $V$ is a finite set of elements referred to as \emph{vertices} and $E\subseteq V^r$ and its elements are referred to as \emph{hyperedges}. Thus, for $r=2$, a hypergraph is simply a digraph, potentially with self-loops. Throughout the text, we will tacitly assume that $V$ is finite. For a hyperedge $h\in E$ we write $v\in h$ if $v$ is an entry of $h$. The degree of a vertex $v\in V$ is defined to be the number of distinct hyperedges $h\in E$ such that $v\in h$, and it is indicated by $\deg(v)$. We say that $\mathbf H$ is \emph{symmetric} if, for every $h\in E$ and every permutation $\pi:[r]\to[r]$, we have that $h^\pi\in E$, where $h^\pi$ is the element of $V^r$ obtained by acting with $\pi$ on the indices of $h$; that is, if $h=(v_1,\ldots,v_r)$, then $h^\pi=(v_{\pi(1)},\ldots,v_{\pi(r)})$.

A path in $\mathbf H$ is a sequence $p=(v_0,h_1,v_1,h_2,\ldots,h_s,v_s)$ where for each $i\in[s]$, it holds that $h_i\in E$ and $v_{i-1},v_i\in h_i$. The length of $p$ is defined to be $s$. A cycle of length $s\ge2$ is a sequence of the same form with $v_s=v_0$, distinct vertices $v_0,\ldots,v_{s-1}$ and distinct hyperedges $h_1,\ldots,h_s$. Similarly to graphs, we define the \emph{girth} of $\mathbf H$ to be the length of the shortest cycle. We say that a hypergraph is \emph{linear} if distinct hyperedges have at most one common entry.

\paragraph{Approximate hypergraph colouring.}
Let $\mathbf H=(V,E)$ and $\mathbf H'=(V',E')$ be $r$-uniform hypergraphs. A map $f:V\to V'$ is a \emph{homomorphism} from $\mathbf H$ to $\mathbf H'$ if for every $h\in E$ with $h=(v_1,\ldots,v_r)$, it holds that $(f(v_1),\ldots,f(v_r))\in E'$. That is, a homomorphism induces a map from the hyperedge relation of one hypergraph to the other. To indicate the existence of a homomorphism from $\mathbf H$ to $\mathbf H'$ we abuse notation and write $\mathbf H\to\mathbf H'$.

Consider the following decision problem: fix an $r$-uniform hypergraph $\mathbf H$, and decide if for an input $r$-uniform hypergraph $\mathbf X$ it holds that $\mathbf X\to\mathbf H$. In the literature of constraint satisfaction, this problem is denoted by $\mathrm{CSP}(\mathbf H)$. For $k,r\ge2$, let $\mathbf H_k^{(r)}=(V_k,E_k^{(r)})$ be the hypergraph whose vertex set is $V_k=\{0,1,\ldots,k-1\}$ and the set of hyperedges is the subset of $V_k^r$ without constant tuples. Namely,
\[
E_k^{(r)}=V_k^r\setminus\{(a,\ldots,a)\in V_k^r\mid a\in V_k\}.
\]
In this formulation, $\mathrm{CSP}(\mathbf H_k^{(r)})$ is the problem of deciding whether an input $r$-uniform hypergraph $\mathbf X$ is $k$-colourable (in the sense that a colour can be assigned to each vertex so that no hyperedge is monochromatic). Note that for $r=3$, $k=2$, this problem is NAE-3SAT where no variable is negated. Indeed, given an instance of monotone \textsf{NAE-3SAT} one can think of it as a $3$-uniform hypergraph where the hyperedges are the clauses and the variables are the vertices. Thus, assigning a Boolean value to each variable so that no clause has only `true' literals is equivalent to $2$-colouring the corresponding hypergraph.

Consider now the following problem: given two fixed $r$-uniform hypergraphs $\mathbf H,\mathbf H'$ such that $\mathbf H\to\mathbf H'$, decide whether, for an input $r$-uniform hypergraph $\mathbf X$, it holds that $\mathbf X\to\mathbf H$ or $\mathbf X\not\to\mathbf H'$ (with the promise that $\mathbf X$ must satisfy exactly one of these conditions). In the literature, this problem is indicated by $\mathrm{PCSP}(\mathbf H,\mathbf H')$, where the first `P' in the acronym stands for \emph{promise}. One can think of $\mathrm{PCSP}(\mathbf H,\mathbf H')$ as the problem of approximating $\mathrm{CSP}(\mathbf H)$ by solving $\mathrm{CSP}(\mathbf H')$, which is a natural problem to consider since the assumption $\mathbf H\to\mathbf H'$ indicates that the constraints in $\mathrm{CSP}(\mathbf H')$ are more relaxed than those of $\mathrm{CSP}(\mathbf H)$.

The hardness result which motivates our reduction concerns the problem of approximating almost balanced hypergraph colouring by hypergraph $3$-colouring. A hypergraph can be almost balanced $2$-coloured if there is a colouring so that, for every hyperedge, the difference between the number of vertices of one colour and the other is at most some small $\epsilon$. Formally, for $r$-uniform hypergraphs and $\epsilon=2$, the problem of deciding whether such a colouring exists is $\mathrm{CSP}(\mathbf B^{(r)})$ where the vertex set of $\mathbf B^{(r)}$ is $\{\pm1\}$ and the set of hyperedges consists of the $r$-tuples whose entries sum to $0$, $\pm1$ or $\pm2$. Approximating this problem for even $r$ by hypergraph $k$-colouring was shown to be NP-hard by Guruswami and Lee. Formally, we have the following.

\begin{theorem}[Guruswami, Lee \cite{GuruswamiLee2018}]\label{thm:gl}
For even values of $r\ge4$ and for all $k\ge3$, the problem $\mathrm{PCSP}(\mathbf B^{(r)},\mathbf H_k^{(r)})$ is NP-hard.
\end{theorem}

\section{Proof of Theorem \ref{thm:main}}

We prove Theorem~\ref{thm:main} by reduction from
$\operatorname{PCSP}\bigl(\mathbf B^{(r)},\mathbf H_3^{(r)}\bigr)$ for even
values of $r$. Specifically, the statement we will prove is the following.

\begin{theorem}\label{thm:fixedarity}
Fix an even $r\ge4$ and let $\alpha_r=1/(r+2)$. Given as input a finite
indexed family $\mathcal V$ of vectors in $\mathbb R^d$ such that
$\|v\|_2^2=\alpha_r$ for every $v\in\mathcal V$ and
\[
  \sum_{v\in\mathcal V}vv^\dagger=I_d,
\]
it is NP-hard to distinguish the following cases:
\begin{enumerate}
    \item $\mathcal V$ such that $W(\mathcal V)=0$.
    \item  $\mathcal{V}$ such that $W(\mathcal V)\ge\beta_r$ where
\[
  \beta_r=\frac1{r+2}\sqrt{r-\frac{4r}{(r-1)^2}}.
\]
\end{enumerate}
\end{theorem}

For the remainder of this section, $r$, $\alpha_r$ and $\beta_r$ will
assume the values in the above statement. We further set
$c=\sqrt{\alpha_r}$ for convenience.

In the very first step of the reduction for the proof of
Theorem~\ref{thm:fixedarity} we use a result by Kun, to reduce this PCSP to the
same problem, but the inputs are restricted to linear hypergraphs.

\begin{theorem}[Kun \cite{Kun2013}]\label{thm:kun}
Let $r,k,g$ be positive integers. For every $r$-uniform hypergraph $\mathbf H$,
there exists a deterministic polynomial-time algorithm that constructs
an $r$-uniform hypergraph $\mathbf H'$ with girth at least $g$, such
that for every $r$-uniform hypergraph $\mathbf T$ with at most $k$ vertices,
it holds that $\mathbf H'\to\mathbf T$ if, and only if,
$\mathbf H\to\mathbf T$.
\end{theorem}

By taking $k=3$ and $g=3$, we can reduce
    $\operatorname{PCSP}\bigl(\mathbf B^{(r)},\mathbf{H}_3^{(r)}$
to considering only instances with girth at least 3. This directly
implies that we can only consider instances of the problem that are linear.
In addition, it also implies that the instances we consider do not have
hyperedges with repeated entries (or these would form a loop and thus imply
that the hypergraph has girth 1).

Hereafter, we fix an even $r$ and consider an $r$-uniform linear
hypergraph $\mathbf H=(V,E)$. Our next step is to construct the coordinate
system in which our Kadison-Singer instance is defined.

\paragraph{Construction of coordinate system.}
We create a coordinate for each hyperedge $h\in E$ and call it $x_h$.
For $v\in V$, let $q_v$ be the smallest power of $2$ at least
$\deg(v)$, and let $j_v=q_v-\deg(v)$. We create a coordinate $x_{v,i}$ for
each $v\in V$ and $i\in[j_v]$ (so, in particular, if $j_v=0$ then there is no
such coordinate). In what follows, it is convenient to denote by $L_{\mathbf H}$
the set of all the labels of our coordinate system; namely,
\[
  L_{\mathbf H}=E\mathbin{\dot\cup}
     \{(v,i)\mid v\in V,i\in[j_v]\}.
\]

\paragraph{Construction of the Kadison-Singer instance.}
Hereafter, we shall work in a real vector space with coordinates $x_l$,
$l\in L_{\mathbf H}$, which we indicate by $\mathbb R^{L_{\mathbf H}}$.
Thus, the matrix that is identity on the coordinates $x_s$, $s\in S$ for some
$S\subseteq L_{\mathbf H}$ and zero on all other coordinates is
indicated by $I_S$. We use $e_t$ for $t\in L_{\mathbf H}$ to denote the
indicator vector of the coordinate $x_t$.
For $v\in V$, define $T_v\subset L_{\mathbf H}$ as
\[
  T_v=\{h\in E\mid v\in h\}\mathbin{\dot\cup}
       \{(v,i)\mid i\in[j_v]\},
\]
noting that $|T_v|=q_v$. Let $Q_v$ be a normalised $T_v\times T_v$ Hadamard
matrix, which is well defined, since we assumed
that $q_v$ is a power of $2$\footnote{Hadamard matrices are well defined for
orders that are powers of $2$. For example, one can take the ordinary $2\times2$
Hadamard matrix
\[
  \begin{pmatrix}1&-1\\1&1\end{pmatrix}
\]
and take its $n$th tensor power to obtain a $2^n\times2^n$ Hadamard
matrix.}. The properties defining $Q_v$, which we will need in the following
construction, are that $Q_vQ_v^\dagger=I_{T_v}$ and the entries of $Q_v$ are
$\pm1/\sqrt{q_v}$.

For each $v\in V$, let
$\iota_v:\mathbb R^{T_v}\to\mathbb R^{L_{\mathbf H}}$ be the inclusion map.
That is, for every vector $w\in\mathbb R^{T_v}$, the vector $\iota_vw$ is
obtained by putting the same entries of $w$ in the entries corresponding to
the coordinates labelled by $T_v$ and $0$ elsewhere. For each $v\in V$ and
$t\in T_v$, we let $w_{v,t}=c\iota_vQ_ve_t$. We call
$w_{v,t}$ a \emph{vertex vector} associated with $v$ and denote the
indexed family of these by $S_v$.

For each $v\in V$, $j\in[j_v]$, $i\in[r+1]$, define the vector
$w_{v,j}^{(i)}\in\mathbb R^{L_{\mathbf H}}$ as $w_{v,j}^{(i)}=ce_{v,j}$.
We refer to these vectors as the \emph{auxiliary indicators} of $v$
and denote the indexed family of these by $A_v$. Finally, for each
$h\in E$ define the vectors $w_{h,1}=w_{h,2}=ce_h$, and refer to these as
the \emph{auxiliary indicators} of $h$ and denote the indexed family
of these by $A_h$. We denote by $\mathcal V_{\mathbf H}$ the indexed
family of all vertex vectors and auxiliary indicators constructed above.
It is easy to see that this family can be constructed in polynomial
time from $\mathbf H$.
In addition, as shown in the next statement, $\mathcal V_{\mathbf H}$
constitutes a valid input to the algorithmic Kadison-Singer problem,
with norm parameter $\alpha_r$.

\begin{lemma}\label{lem:validity}
If $\mathcal V_{\mathbf H}$ as above is non-empty, then for every
$w\in\mathcal V_{\mathbf H}$ it holds that
$\|w\|_2^{2}=\alpha_r$ and
\[
  \sum_{w\in\mathcal V_{\mathbf H}}ww^\dagger=I_{L_{\mathbf H}}.
\]
\end{lemma}

\begin{proof}
If $w$ is an auxiliary indicator, it follows from the definition that
$\|w\|_2^{2}=w^\dagger w=\alpha_r$. For a vertex vector $w_{v,t}$ it
holds that
\begin{equation}\label{eq:1}
  w_{v,t}^\dagger w_{v,t}
    =c^2e_t^\dagger Q_v^\dagger\iota_v^\dagger\iota_vQ_ve_t.
\end{equation}
Noting that $\iota_v^\dagger\iota_v=I_{T_v}$ and that
$Q_v^\dagger Q_v=I_{T_v}$ by the properties of normalised Hadamard matrices,
it then holds that $\|w_{v,t}\|_2^{2}=c^2=\alpha_r$ as required.
Consider now the following matrix:
\begin{equation}\label{eq:2}
  \sum_{v\in V}\sum_{w\in S_v}ww^\dagger.
\end{equation}
It is diagonal by construction. Since $T_v\cap T_{v'}\subseteq E$ for
distinct $v,v'$, it must hold that the $x_l$ diagonal entries of the matrix
in Equation~\eqref{eq:2} for $l\in T_v\setminus E$ must be equal to
$\alpha_r$. This is because, for a fixed $v\in V$,
$\sum_{w\in S_v}ww^\dagger=\alpha_rI_{T_v}$. In addition, the $x_l$ diagonal
entries for $l\in E$ must be equal to $r\alpha_r$, since each hyperedge $h$
occurs in exactly $r$ distinct $T_v$ because its $r$ entries are
distinct by linearity of $\mathbf{H}$. Note that
\begin{equation}\label{eq:3}
  \sum_{w\in A_v}ww^\dagger
    =\sum_{j\in[j_v]}\sum_{i\in[r+1]}c^2e_{v,j}e_{v,j}^\dagger
    =c^2(r+1)I_{T_v\setminus E}
    =\frac{r+1}{r+2}I_{T_v\setminus E}.
\end{equation}
In addition, we have
\begin{equation}\label{eq:4}
  \sum_{w\in A_h}ww^\dagger
    =2c^2I_{\{h\}}=2\alpha_rI_{\{h\}}
    =\frac2{r+2}I_{\{h\}}.
\end{equation}
since $A_h$ contains exactly two copies of the vectors $ce_h$ for each
$h\in E$. By summing Equations~\eqref{eq:2}, \eqref{eq:3}, \eqref{eq:4}
over the relevant vertices and edges, each coordinate receives
$(r+2)\alpha_r=1$, thus yielding the identity as required.
\end{proof}

Now we get to the meat of the hardness proof. First, we show that if
$\mathbf H$ is almost balanced colourable, then
$W(\mathcal V_{\mathbf H})=0$.

\begin{lemma}\label{lem:completeness}
Let $\mathbf H$ be an $r$-uniform hypergraph as above for which
$\mathbf H\to\mathbf B^{(r)}$. Then $W(\mathcal V_{\mathbf H})=0$.
\end{lemma}

\begin{proof}
Let $f$ be a homomorphism from $\mathbf H$ to $\mathbf B^{(r)}$. Note that
since $r$ is even, it must be that for every hyperedge
$h=(v_1,\ldots,v_r)\in E$, it must hold that
$b_h:=\sum_{i\in[r]}f(v_i)\in\{0,\pm2\}$. Define a map
$\sigma:\mathcal V_{\mathbf H}\to\{\pm1\}$ as follows. For every $v\in V$
and $w\in S_v$ define $\sigma(w)=f(v)$. For every $v\in V$
and $j\in[j_v]$, we let $\sigma$ evaluate to $-f(v)$ for $r/2+1$ of
the auxiliary indicators $w_{v,j}^{(i)}$, and $f(v)$ for the remaining
$r/2$. Since there are $r+1$ of these vectors for each
private coordinate $(v,j)$, we have that
$\sum_{i\in[r+1]}\sigma(w_{v,j}^{(i)})=-f(v)$.
For each $h\in E$, recall that there are only two auxiliary indicators
$w_{h,1}$ and $w_{h,2}$. Since $b_h\in\{0,\pm2\}$, it follows that we can
choose the value of $\sigma$ on these auxiliary indicators so that
$\sigma(w_{h,1})+\sigma(w_{h,2})=-b_h$.

We now show that for this choice of $\sigma$ it holds that
$W_\sigma(\mathcal V_{\mathbf H})=0$. The matrix
\begin{equation}\label{eq:5}
  \sum_{w\in\mathcal V_{\mathbf H}}\sigma(w)ww^\dagger
\end{equation}
is diagonal. This is because $\sigma$ is constant on the vertex vectors
in $S_v$ for each $v\in V$, and the auxiliary indicators are all single
entry vectors. In particular, for each $v\in V$,
\begin{equation}\label{eq:6}
  \sum_{w\in S_v}\sigma(w)ww^\dagger=f(v)\alpha_rI_{T_v}.
\end{equation}
We also have
\begin{equation}\label{eq:7}
\begin{aligned}
  \sum_{w\in A_v}\sigma(w)ww^\dagger
    &=\alpha_r\sum_{j\in[j_v]}
       \left(\sum_{i\in[r+1]}\sigma(w_{v,j}^{(i)})\right)I_{\{(v,j)\}}\\
    &=-f(v)\alpha_rI_{T_v\setminus E}
\end{aligned}
\end{equation}
and
\begin{equation}\label{eq:8}
  \sum_{w\in A_h}\sigma(w)ww^\dagger=-b_h\alpha_rI_{\{h\}}.
\end{equation}
By summing Equations~\eqref{eq:6} and \eqref{eq:7} over $v\in V$ we
obtain a diagonal matrix whose only non-zero diagonal entries
correspond to the $x_h$ coordinates for $h\in E$. For each $h$, these entries
are equal to $\alpha_r\sum_{v\in h}f(v)=\alpha_rb_h$ and thus,
adding Equation~\eqref{eq:8} over $h\in E$ gives a $0$-matrix as
required.
\end{proof}


To conclude the proof of Theorem~\ref{thm:fixedarity} we prove the following.

\begin{lemma}\label{lem:soundness}
Let $\mathbf H$ be an $r$-uniform linear hypergraph such that
$\mathbf H\not\to\mathbf H_3^{(r)}$. Then $W(\mathcal V_{\mathbf H})\ge\beta_r$, where
$\beta_r$ is as in Theorem~\ref{thm:fixedarity}.
\end{lemma}

In order to show the above statement we first need to analyse $W_\sigma(\mathcal V_{\mathbf H})$
for arbitrary $\sigma$ through two auxiliary lemmas. First let us fix some
$\sigma:\mathcal V_{\mathbf H}\to\{\pm1\}$ and define
\[
 Z=\sum_{w\in\mathcal V_{\mathbf H}}\sigma(w)ww^\dagger.
\]
For each $v\in V$, let $a_v$ be the average sign over all vertex vectors:
\begin{equation}\label{eq:9}
 a_v=\frac1{q_v}\sum_{w\in S_v}\sigma(w).
\end{equation}
Further define the matrix
\[
 M_v=Q_vD_{v,\sigma}Q_v^\dagger
\]
where $D_{v,\sigma}$ is the $T_v\times T_v$ diagonal matrix whose
$(l,l)$-entry is $\sigma(w_{v,l})$ for $l\in T_v$. By construction,
$M_v$ is symmetric and orthogonal and thus, $M_v^2=I_{T_v}$. In addition,
notice that
\[
 (M_v)_{jj}=\sum_{t\in T_v}(Q_v)_{jt}^2\sigma(w_{v,t})
 =\sum_{t\in T_v}\frac1{q_v}\sigma(w_{v,t})=a_v.
\]
Thus, since $M_v$ is orthogonal, we must have that the sum of the squares of
the non-diagonal entries of a column (or a row) must equal $1-a_v^2$.
We shall abuse language and invoke this property later in this paper by
referring to it as `orthogonality of $M_v$'. The next Lemma shows a lower
bound on $\|Z\|_2$, which we shall use to obtain a lower bound on
$W(\mathcal V_{\mathbf H})$.

\begin{lemma}\label{lem:column}
Let $\tau_h=\sigma(w_{h,1})+\sigma(w_{h,2})$. For every $h\in E$, it holds that
\[
 \|Ze_h\|_2^2
 =\alpha_r^2\left[\left(\sum_{v\in h}a_v+\tau_h\right)^2
                  +r-\sum_{v\in h}a_v^2\right].
\]
\end{lemma}

\begin{proof}
The vector $Ze_h$ is equal to the $h$-column of $Z$. This means that it is equal
to the $h$-column of the matrix
\begin{equation}\label{eq:10}
 \sum_{v\in h}\sum_{w\in S_v}\sigma(w)ww^\dagger
 +\sum_{w\in A_h}\sigma(w)ww^\dagger.
\end{equation}
The second term of the above equation only contributes to the $h$-entry of
$Ze_h$. In particular, it contributes $\alpha_r\tau_h$. Let us rewrite the
first term as follows:
\begin{equation}\label{eq:11}
 \sum_{v\in h}\sum_{w\in S_v}\sigma(w)ww^\dagger
 =\sum_{v\in h}\alpha_r\iota_vQ_vD_{v,\sigma}Q_v^\dagger
                         \iota_v^\dagger
 =\sum_{v\in h}\alpha_r\iota_vM_v\iota_v^\dagger.
\end{equation}
To check this, it is sufficient to verify the action of the above operators
on the unit vectors $e_t$, $t\in L_{\mathbf H}$.
For distinct $u,v\in h$, linearity gives $T_u\cap T_v=\{h\}$.
Consequently, after removing their $h$-entries, the $h$-columns of the
embedded matrices $\iota_vM_v\iota_v^\dagger$, $v\in h$, have pairwise
disjoint supports. Their squared norms therefore add without the diagonals interfering.
The sum of the squares of the off-diagonal entries of the $h$-column
in the above is therefore equal to
\begin{equation}\label{eq:12}
 \alpha_r^2\sum_{v\in h}(1-a_v^2)
 =\alpha_r^2\left(r-\sum_{v\in h}a_v^2\right)
\end{equation}
by the orthogonality of $M_v$ and the disjointness of these supports.
The $h$-diagonal entry of the matrix in Equation~\eqref{eq:11} is equal
to $\alpha_r\sum_{v\in h}a_v$ using the property of $M_v$ and thus, the
$h$-entry of $Ze_h$ is therefore
\begin{equation}\label{eq:13}
 \alpha_r\left(\sum_{v\in h}a_v+\tau_h\right).
\end{equation}
Combining this with Equation~\eqref{eq:12} yields the statement of the Lemma.
\end{proof}

The next statement gives the required lower bound for $\|Ze_h\|_2^2$
which translates to a lower bound for $W(\mathcal V_{\mathbf H})$. In what follows, we set
\begin{equation}\label{eq:14}
 \kappa=\frac2{r-1}
\end{equation}
and
\[
 C_r=r-\frac{4r}{(r-1)^2}=r(1-\kappa^2).
\]
We assume that $r\ge4$ so that $\kappa\in(0,1)$ and $C_r>0$.
Let $y_1,\ldots,y_r\in[-1,1]$ be constants and $\tau\in\{0,\pm2\}$ and define
\[
 F_r(y_1,\ldots,y_r,\tau)
 =\left(\sum_{i\in[r]}y_i+\tau\right)^2+r-\sum_{i\in[r]}y_i^2.
\]

\begin{lemma}\label{lem:scalar}
If $y_1,\ldots,y_r$ all lie in one of the intervals $[-1,-\kappa]$,
$[-\kappa,\kappa]$, $[\kappa,1]$ then
\[
 F_r(y_1,\ldots,y_r,\tau)\ge C_r
\]
for any $\tau\in\{0,\pm2\}$.
\end{lemma}

\begin{proof}
Let us first consider the interval $[-\kappa,\kappa]$, so assume $|y_i|\le\kappa$
for all $i\in[r]$. Then
\[
 \sum_{i\in[r]}y_i^2\le r\kappa^2.
\]
From this, we get that $F_r\ge r-r\kappa^2=C_r$ as required.

Now consider the interval $[\kappa,1]$, so assume $y_i\ge\kappa$ for all
$i\in[r]$. Then
\[
 \sum_{i\in[r]}y_i\ge r\kappa=\frac{2r}{r-1}>2.
\]
Thus, by picking $\tau=-2$, we minimise the value of the square term in $F_r$,
so let us assume $\tau=-2$ from now on. Let $b_i=y_i-\kappa$ for $i\in[r]$
(in particular, $b_i\ge0$) and set
\[
 B=\sum_{i\in[r]}b_i.
\]
By noticing that $r\kappa-2=\kappa$ we may bound $F_r$ from below as follows
\begin{align}
 F_r(y_1,\ldots,y_r,-2)
  &=(r\kappa+B-2)^2+r-r\kappa^2-2\kappa B-\sum_{i\in[r]}b_i^2
       \label{eq:15}\\
  &=r-(r-1)\kappa^2{}+{}B^2-\sum_{i\in[r]}b_i^2
       \label{eq:16}\\
  &=r-\frac4{r-1}+2\sum_{1\le i<j\le r}b_ib_j
       \label{eq:17}\\
  &\ge r-\frac4{r-1}.
       \label{eq:18}
\end{align}
Finally,
\[
 r-\frac4{r-1}=C_r+\frac4{(r-1)^2}\ge C_r
\]
as required.

For the interval $[-1,-\kappa]$ the argument is the same as the above, except
that we replace $y_i$ by $-y_i$ and $\tau$ by $-\tau$.
\end{proof}

We may finally prove Lemma~\ref{lem:soundness}.

\begin{proof}[Proof of Lemma~\ref{lem:soundness}]
Assume that $\mathbf H$ is not $3$-colourable. Hence, for every map
$f:V\to\{0,1,2\}$, some $h\in E$ must be monochromatic; that is, there is
some $\chi\in\{0,1,2\}$ such that $f(v)=\chi$ for all $v\in h$.
Consider a map $f_\sigma$ colouring the vertices of $\mathbf H$ by their average
sign $a_v$ for some $\sigma:\mathcal V_{\mathbf H}\to\{\pm1\}$ in the sense of
Equation~\eqref{eq:9}. Namely,
\[
 f_\sigma(v)=
 \begin{cases}
 0&\text{if }a_v<-\kappa,\\
 1&\text{if }|a_v|\le\kappa,\\
 2&\text{if }a_v>\kappa,
 \end{cases}
\]
where $\kappa$ is as in Equation~\eqref{eq:14}. Since $f_\sigma$ is not a
$3$-colouring, there must be some $h\in E$ for which $a_v$ lie in one of the
intervals $[-1,-\kappa]$, $[-\kappa,\kappa]$, $[\kappa,1]$ for all
$v\in h$. Thus, for any $\sigma:\mathcal V_{\mathbf H}\to\{\pm1\}$, it must hold that
\begin{equation}\label{eq:19}
 W_\sigma(\mathcal V_{\mathbf H})^2=\|Z\|_2^2
 \ge\|Ze_h\|_2^2\ge\alpha_r^2C_r=\beta_r^2
\end{equation}
where the last inequality follows by combining Lemmas~\ref{lem:column}
and~\ref{lem:scalar}.
\end{proof}

\begin{proof}[Proof of Theorem~\ref{thm:fixedarity}]
Combine Lemmas \ref{lem:completeness} and~\ref{lem:soundness}.
\end{proof}

\begin{proof}[Proof of Theorem~\ref{thm:main}]
Note that
\[
 \frac{\beta_r}{\sqrt{\alpha_r}}
 =\sqrt{\frac r{r+2}\left(1-\frac4{(r-1)^2}\right)}
\]
which asymptotically tends to $1$ as $r\to\infty$ and it is never larger than $1$.

\end{proof}
\section{Conclusion}
The reader may note that we talked about an indexed family of vectors as the input to $\mathsf{KS}_2(c)$, rather than a set of vectors. This is because we are allowed to take multiple copies of the same vector as it can be seen in our hardness reduction as well as the one by Spielman and Zhang \cite{SpielmanZhang2022}. It would be interesting to show whether the same hardness bound holds if the input is not allowed to have multiple copies of the same vector. Jourdan, Macgregor and Sun \cite{JourdanMacgregorSun2023} achieve hardness for $c=1/(2\sqrt{2})$ using instances where such occurrences are not allowed.

A easy exercise shows that if $m$ is the number of vectors in our input and $d$ is the dimension of the vector space, and if $r$ is fixed, then for the instances of $\mathsf{KS}_2(\alpha_r)$ constructed in our reduction it holds that $m=\Theta(d)$. Hence, we do not get any `dense' instances in the sense of \cite{JourdanMacgregorSun2024Dense}.

The question motivating this note was, in light of recent research, to estimate the supremum $c_K$ of all constants $c$ for which $\mathsf{KS}_2(c)$ is \textsf{NP}-hard. It is unclear if our main result shows that $c_K=1$, but this suggests investigating the problem $\mathsf{KS}_2(1)$.

\paragraph{Acknowledgements.} I would like to thank Ben Jourdan, Peter Macgregor and He Sun for introducing me to the problem and for sharing insightful discussions. I am grateful to Lorenzo Ciardo for suggesting the $d$-to-$1$ Games hardness direction at CWC'25 in Cervinia.

\paragraph{Statement of AI use.} The author originally proved hardness of $\mathsf{KS_2}(\sqrt{5/32})$ using PCSP$\left(\mathbf{H}_2^{(3)},\mathbf{H}_3^{(3)}\right)$ as source of hardness. For this, ChatGPT was used to solve an optimisation problem similar to the one in Lemma \ref{lem:scalar}. This hardness proof was used as a template to show Theorem \ref{thm:main} conditional on the Rich $2$-to-$1$ with perfect completeness conjecture, using a result by Banakh, Barsukov and Nakajima \cite{bbn}. Whilst proofreading the original manuscript with ChatGPT, it was suggested to the author that essentially the same proof yielded Theorem \ref{thm:main} unconditionally, through the use of Theorem \ref{thm:gl} which was unknown to the author. The author accepts full responsibility for the text.
\begingroup
\bibliographystyle{plain}
\bibliography{weaver_references}
\endgroup

\end{document}